\documentclass[12pt]{article}
\usepackage{graphicx}
\usepackage{lscape}
\begin{document}

\def\ds{\displaystyle}
\def\beq{\begin{equation}}
\def\eeq{\end{equation}}
\def\bea{\begin{eqnarray}}
\def\eea{\end{eqnarray}}
\def\ve{\vert}
\def\vel{\left|}
\def\ver{\right|}
\def\nnb{\nonumber}
\def\ga{\left(}
\def\dr{\right)}
\def\aga{\left\{}
\def\adr{\right\}}
\def\lla{\left<}
\def\rra{\right>}
\def\rar{\rightarrow}
\def\nnb{\nonumber}
\def\la{\langle}
\def\ra{\rangle}
\def\ba{\begin{array}}
\def\ea{\end{array}}
\def\tr{\mbox{Tr}}
\def\ssp{{\Sigma^{*+}}}
\def\sso{{\Sigma^{*0}}}
\def\ssm{{\Sigma^{*-}}}
\def\xis0{{\Xi^{*0}}}
\def\xism{{\Xi^{*-}}}
\def\qs{\la \bar s s \ra}
\def\qu{\la \bar u u \ra}
\def\qd{\la \bar d d \ra}
\def\qq{\la \bar q q \ra}
\def\gGgG{\la g^2 G^2 \ra}
\def\q{\gamma_5 \not\!q}
\def\x{\gamma_5 \not\!x}
\def\g5{\gamma_5}
\def\sb{S_Q^{cf}}
\def\sd{S_d^{be}}
\def\su{S_u^{ad}}
\def\ss{S_s^{??}}
\def\sbp{{S}_Q^{'cf}}
\def\sdp{{S}_d^{'be}}
\def\sup{{S}_u^{'ad}}
\def\ssp{{S}_s^{'??}}
\def\sig{\sigma_{\mu \nu} \gamma_5 p^\mu q^\nu}
\def\fo{f_0(\frac{s_0}{M^2})}
\def\ffi{f_1(\frac{s_0}{M^2})}
\def\fii{f_2(\frac{s_0}{M^2})}
\def\O{{\cal O}}
\def\sl{{\Sigma^0 \Lambda}}
\def\es{\!\!\! &=& \!\!\!}
\def\ar{&+& \!\!\!}
\def\ek{&-& \!\!\!}
\def\cp{&\times& \!\!\!}
\def\se{\!\!\! &\simeq& \!\!\!}
\title{
         {\Large
                 {\bf
Octet baryon magnetic moments in light cone QCD sum rules
                 }
         }
      }

\author{\vspace{1cm}\\
{\small T. M. Aliev$^a$ \thanks
{e-mail: taliev@metu.edu.tr}\,\,,
A. \"{O}zpineci$^b$ \thanks
{e-mail: ozpineci@ictp.trieste.it}\,\,,
M. Savc{\i}$^a$ \thanks
{e-mail: savci@metu.edu.tr}} \\
{\small a Physics Department, Middle East Technical University, 
06531 Ankara, Turkey}\\
{\small b  The Abdus Salam International Center for Theoretical Physics,
I-34100, Trieste, Italy} }
\date{}

\begin{titlepage}
\maketitle
\thispagestyle{empty}

\begin{abstract}
Octet baryon magnetic moments are calculated in framework of the
light cone QCD sum rules. The analysis is carried for the general form 
of the interpolating currents for octet baryons. A comparison of our 
results on the magnetic moments of octet baryons with the predictions of
other approaches and experimental data is presented.  
\end{abstract}

~~~PACS number(s): 11.55.Hx, 13.40.Em, 14.20.Jn
\end{titlepage}

\section{Introduction}

The method of QCD sum rule \cite{R1} has become a powerful tool in studying
hadron physics on the basis of QCD. It is a framework which connects
physical parameters of hadrons to the QCD parameters. In this approach
hadrons are represented by their interpolating quark currents taken at large
virtualities and then correlation function of these quark currents is 
introduced. Following this, on one side, the correlation function is
calculated by the operator product expansion (OPE), and on the other side, 
its phenomenological part is constructed. Physical quantities of interest are
determined by matching these two descriptions at large virtualities via
dispersion relations.

The QCD sum rule method is discussed in many review articles (see
\cite{R2}--\cite{R6} and references therein) and successfully applied in
studying various characteristics of hadrons. One of the important 
characteristic static parameters of hadrons is their magnetic moments. Magnetic
moments of nucleon were calculated in framework of the QCD sum rule approach 
in \cite{R7,R8} using external field technique. They were later refined and
extended to the entire baryon octet \cite{R9,R10}.

In this work, we present an independent calculation of the magnetic
moments of the octet baryons in the framework of an alternative approach to
the traditional sum rules method, i.e., the light cone QCD sum rules method
(LCQSR). In the LCQSR method OPE is carried out near the light cone $x^2=0$
instead of at the short distance $x\approx 0$. This is an expansion over the
twists of the operators rather than the dimensions, as is the case in the
traditional QCD sum rules. The nonperturbative dynamics is parametrized
by the so called light cone wave functions, rather than the vacuum
condensates used in the traditional QCD sum rules. A detailed description of
this method can be found in \cite{R6,R11,R12}. There are many applications
of the LCQSR in the current literature. Note that magnetic moments of the
nucleon and decuplet baryons were studied in \cite{R13} and
\cite{R14},respectively, using the LCQSR method.

The paper is organized as follows. In section 2 the LCQSR for the magnetic
moments of the octet baryons are derived. In section 3 we present our
numerical calculations and a comparison of the prediction of the LCQSR
method on the magnetic moments of the octet baryons with the results of the
other methods and experimental results.

\section{LCQSR for the magnetic moments of the octet baryons}

As we have already noted, in order to construct the sum rules for the
magnetic moments of hadrons, it is necessary to introduce the correlation
function of interpolating quark currents which have the same quantum numbers
as corresponding hadrons. For this aim we consider the following correlator
function:
\bea
\label{e1}
\Pi = i \int d^4 x \, e^{i p x} \lla 0 \ve T\{\eta_B(x)
\bar \eta_B(0) \} \ve 0 \rra_\gamma~,
\eea
where $\gamma$ means the external electromagnetic field, $\eta_B$ is the 
interpolating current of the corresponding baryon. It follows from this
expression that we need the explicit expression of the interpolating currents
to calculate the correlation function. It is well known that there is a
continuum number of interpolating currents for the baryons. The 
general form of the interpolating currents for the octet
baryons can be represented as \cite{R15,R16}
\bea  
\label{e2}
\eta_p \es 2 \epsilon_{abc} \sum_{\ell=1}^2 
\ga u^{aT} C A_1^\ell d^B \dr A_2^\ell u^c~,\nnb \\
\eta_n \es n_p \ga u^c \rar d^c \dr~,\nnb \\
\eta_{\Sigma^0} \es \frac{1}{\sqrt{2}} \epsilon_{abc} \sum_{\ell=1}^2
\Big[\ga u^{aT} C A_1^\ell s^b\dr A_2^\ell d^c -
\ga d^{cT} C A_1^\ell s^b\dr A_2^\ell u^a\Big]~,\nnb \\
\eta_{\Lambda} \es \frac{2}{\sqrt{6}}\epsilon_{abc} \sum_{\ell=1}^2
\Big[-2\ga u^{aT} C A_1^\ell d^b\dr A_2^\ell s^c +
\ga u^{aT} C A_1^\ell s^b\dr A_2^\ell d^c +
\ga d^{cT} C A_1^\ell s^b\dr A_2^\ell u^a \Big]~,\nnb \\
\eta_{\Xi^0} \es -2 \epsilon_{abc} \sum_{\ell=1}^2
\ga s^{aT} C A_1^\ell u^b\dr A_2^\ell s^c~,\nnb \\
\eta_{\Sigma^+} \es \eta_{\Sigma^0} \ga d \rar u \dr~,\nnb \\
\eta_{\Sigma^-} \es \eta_{\Sigma^0} \ga u \rar d \dr~,\nnb \\
\eta_{\Xi^-} \es \eta_{\Xi^0} \ga u \rar d \dr~,
\eea
where $a,b,c$ are the color indices, $A_1^1 = 1$,  $A_1^2 = t \gamma_5$,
$A_2^1 = \gamma_5$, $A_2^2 = t$ and $t$ is an arbitrary parameter.
Ioffe current corresponds to the choice $t=-1$. 

Let us firstly discuss the hadronic representation of the correlator function. 
By inserting a complete set of states between the currents in 
Eq. (\ref{e1}) with quantum numbers of the corresponding baryon $B$, one
obtains the hadronic representation of the correlator
\bea
\label{e3}
\Pi \es \frac{\lla 0 \ve \eta_{B} \ve B(p_1) \rra}
{p_1^2 - m_{B}^2} \lla B(p_1) \ve B (p_2)
\rra_\gamma
\frac{\lla B(p_2) \ve \bar \eta_{B} \ve 0 \rra}{p_2^2 -
m_{B}^2} \nnb \\
\ar \sum_{h}
\frac{\lla 0 \ve \eta_{B} \ve h(p_1) \rra}
{p_1^2 - m_{h}^2} \lla h(p_1) \ve h (p_2) \rra_\gamma
\frac{\lla h (p_2) \ve \bar \eta_{B} \ve 0 \rra}{p_2^2 -
m_{h}^2}~,
\eea
where $p_2 = p_1 + q$, and $q$ is the photon momentum. The second term in
Eq. (\ref{e3}) takes into account higher states and continuum contributions
and $h$ forms a complete set of baryons having the same quantum number as
the ground state baryon $B$ (here $B$ represents any one of
$p,~n,~\Sigma,~\Lambda$ and $\Xi$). The coupling strength of the
interpolating currents with the baryon states parametrized with the overlap
amplitudes $\lambda_B$ is defined as
\bea
\label{e4}
\lla 0 \ve \eta_{B} \ve B(p) \rra = \lambda_B u_B(p)~.
\eea
It follows from Eq. (\ref{e3}) that in order to obtain an expression for the
phenomenological part of the correlator, explicit form of the matrix element 
$\lla B(p_1) \ve B (p_2) \rra_\gamma$ is needed. Electromagnetic vertex of
the spin $1/2$ baryons can be written as
\bea
\label{e5}
\lla B(p_1) \ve B(p_2) \rra_\gamma \es
\bar u_B(p_1) \left[ f_1 \gamma_\mu + \frac{i \sigma_{\mu\nu} q^\nu}
{2 m_B} f_2 \right] u_B(p_2) \varepsilon^\mu~, \nnb \\
\es \bar u_B(p_1) \left[ (f_1 + f_2) \gamma_\mu + \frac{(p_1 + p_2)_\mu}
{2 m_B} f_2 \right] u_B(p_2) \varepsilon^\mu~,
\eea
where the form factors $f_i$ are in general 
functions of $q^2 = (p_2 - p_1)^2$, and $\varepsilon^\mu$ is the
polarization vector of the photon. In the problem under consideration, only the
values of the form factors at $q^2=0$ are needed.

Using Eqs. (\ref{e3})--(\ref{e5}), for the phenomenological part of the
correlator (\ref{e1}) we obtain
\bea
\label{e6}
\Pi  = - \lambda_B^2 \varepsilon^\mu \frac{\not\!p_1 + m_B}{p_1^2 - m_B^2}
\left[ (f_1 + f_2) \gamma_\mu + \frac{(p_1 + p_2)_\mu}{2 m_B}
f_2 \right] \frac{\not\!p_2 + m_B}{p_2^2 - m_B^2} + \cdots~,
\eea
where $\cdots$ represents contributions from the higher states and the
continuum. As is obvious from Eq. (\ref{e6}), the correlator function
contains some number of different structures. Among all possible structures,
we choose the one $\sim \not\!p_1 \not\!\varepsilon \not\!q$ that contains
the magnetic form factor $f_1+f_2$. The value of $f_1+f_2$ at $q^2=0$ gives
the magnetic moment of the baryon in units of its natural magneton, i.e., 
$e\hbar/2 m_B c$. 

Isolating the structure $\sim \not\!p_1 \not\!\varepsilon \not\!q$ from the
phenomenological part of the correlator, which describes the magnetic form
factor, we get
\bea
\label{e7}
\Pi = - \lambda_B^2 \frac{1}{p_1^2-m_B^2} \mu \frac{1}{p_2^2-m_B^2} + \cdots~,
\eea
where $\mu = \ga f_1 + f_2 \dr \ve_{q^2=0}$.

Using the explicit form of the interpolating currents given in Eq.
(\ref{e2}) and after some calculations, for the QCD part of the correlator
functions of the $\Lambda$ and $\Sigma^+$ baryons, we get
\bea
\label{e8}
\lefteqn{
\Pi^\Lambda =
- \frac{2}{3} \epsilon_{a b c} \epsilon_{d e f} \int d^4x e^{i p x} 
\la \gamma \ve \sum_{\ell=1}^2 \, \sum_{k=1}^2 \Big\{
4 A_2^\ell S_s^{be}(x) A_2^k \, \tr S_d^{cf}(x) C A_1^k S_u^{adT}(x) 
C A_1^\ell} \nnb \\
\ar 2 A_2^\ell S_s^{be}(x) C A_1^k S_u^{adT}(x) C A_1^\ell S_d^{cf}(x) 
A_2^k
- 2 A_2^\ell S_s^{be}(x) C A_1^k S_d^{cfT}(x)(C A_1^\ell)^T S_u^{ad}(x) 
A_2^k \nnb \\
\ar 2 A_2^\ell S_d^{cf}(x) C A_1^k S_u^{adT}(x) C A_1^\ell S_s^{be}(x) 
A_2^k
+ A_2^\ell S_d^{cf}(x) A_2^k \, \tr S_s^{be}(x) C A_1^k S_u^{adT}(x) 
C A_1^\ell \\
\ek A_2^\ell S_d^{cf}(x) (C A_1^k)^T S_s^{beT}(x) (C A_1^\ell)^T 
S_u^{ad}(x) A_2^k
- 2 A_2^\ell S_u^{ad}(x) (C A_1^k)^T S_d^{cfT}(x) C A_1^\ell 
S_s^{be}(x) A_2^k \nnb \\
\ek A_2^\ell S_u^{ad}(x) (C A_1^k)^T S_s^{beT}(x) (C A_1^\ell)^T 
S_d^{cf}(x) A_2^k
+ A_2^\ell S_u^{ad}(x) A_2^k \, \tr S_s^{be}(x) C A_1^k S_d^{cfT}(x) 
C A_1^\ell \Big\} \ve 0 \ra~, \nnb \\ \nnb \\ \nnb
\label{e9}
\lefteqn{
\Pi^{\Sigma^+} = -2  \epsilon_{a b c} \epsilon_{d e f} \int d^4x e^{i p x}
\la \gamma \ve \sum_{\ell=1}^2 \, \sum_{k=1}^2 \Big\{
A_2^\ell S_u^{cf}(x) A_2^k \, \tr S_s^{be}(x) C A_1^k S_u^{adT}(x) C
A_1^\ell} \nnb \\
\ar A_2^\ell \Big[ S_u^{cf}(x) (C A_1^k)^T S_s^{beT}(x) (C A_1^\ell)^T
S_u^{ad}(x) A_2^k 
+ S_u^{ad}(x) (C A_1^k)^T S_s^{beT}(x) (C A_1^\ell)^T 
S_u^{cf}(x) A_2^k \nnb \\
\ar S_u^{ad}(x) A_2^k \, \tr S_s^{be}(x) C A_1^k S_d^{cfT}(x) C A_1^\ell
\Big] \Big\} \ve 0 \ra~,
\eea 
where $C$ is the charge conjugation operator, subscripts $a,b,c,d,e,f$
are the color indices and $S_q$ is the full propagator of the light quark
involving both perturbative and nonperturbative parts. The expression for 
the theoretical parts of the correlator functions of the $\Sigma^-,~\Sigma^0,~
\Xi^0$ and $\Xi^-$ baryons can be obtained from Eq. (\ref{e9}) by the
following replacements:
\bea
\label{e10}
\Pi^{\Sigma^-} \es \Pi^{\Sigma^+} (u \rar d)~,\nnb \\
\Pi^{\Xi^0} \es \Pi^{\Sigma^+} (u \leftrightarrow s)~,\nnb \\
\Pi^{\Xi^-} \es \Pi^{\Sigma^+} (u \rar s,~s\rar d)~,\nnb \\
\Pi^{\Sigma^0} \es \frac{1}{2} \ga \Pi^{\Sigma^-} + \Pi^{\Sigma^+} \dr~,\nnb \\
\Pi^p \es \Pi^{\Xi^-} (s\rar u)~.
\eea
Our calculations show that contributions 
from quadratic terms in strange quark mass $m_s$ are negligibly small compared 
to that of linear terms in $m_s$, which is about $6\%$. Because of this reason
theoretical part of the correlator function is calculated up to linear order in
$m_s$. For the full light quark propagator we have used the following
expression
\bea
\label{e11}
\lefteqn{
S_q(x) = \lla 0 \vel T \left\{ \bar q(x) q(0) \right\} \ver 0 \rra}\nnb \\
\es \frac{i \not\!x}{2 \pi^2 x^4} - \frac{m_q}{4 \pi^2 x^2} - \frac{\qq}{12}
\ga 1 - \frac{i m_q}{4} \not\!x \dr - \frac{x^2}{192} m_0^2 \qq 
\ga 1 - \frac{i m_q}{6} \not\!x \dr  \\
\ek i g_s \int_0^1 dv \Bigg[ \frac{\not\!x}{16 \pi^2 x^2} G_{\mu\nu}(vx)
\sigma^{\mu\nu} - v x^\mu G_{\mu\nu} (vx) \gamma^\nu \frac{i}{4 \pi^2 x^2}
- \frac{i m_q}{32 \pi^2}  G_{\mu\nu} \sigma^{\mu\nu}
\ga \ln \frac{-x^2 \Lambda^2}{4} + 2 \gamma_E \dr \Bigg]~,\nnb
\eea
where $\Lambda$ is the energy cut off separating perturbative and
nonperturbative regimes.

Few words about this expression of the light quark propagator are in order.
The complete light cone expansion of the light quark propagator in external
field has been carried in \cite{R17}, and it has been shown that it gets
contributions from nonlocal quark operators $\bar q G q,~\bar q G G q,~\bar
q q \bar q q$, where $G_{\mu\nu}$ is the gluon field strength tensor. In the
present work we consider the operators with only one--gluon field and neglect
components with two--gluon and four--quark field. Formally, neglecting 
two--gluon and four--quark field terms can be justified on the basis of an
expansion in conformal spin \cite{R18}.  

Perturbative part (i.e., photon interacting with quarks perturbatively)
of the correlator function can be obtained by making the following substitution
in one of the propagators in Eq. (\ref{e9})
\bea
\label{e12}
{S_q}_{\alpha\beta}^{ab} \rar -\frac{1}{2} e e_q \ga \int dy \, {\cal F}^{\mu\nu} y_\nu 
S_q^{free} (x-y) \gamma_\mu S_q^{free} (y) \dr_{\alpha\beta}^{ab}~,
\eea
where the Fock--Schwinger gauge $x^\mu A_\mu(x)=0$ is used and 
$S_q^{free}=i\not\!x/(2\pi^2x^4)$ is the free propagator, the remaining two
propagators are the full propagators of the quarks (see Eq. (\ref{e9})) and 
${\cal F}^{\mu\nu}$ is the electromagnetic field strength tensor. The explicit
expression of the nonperturbative contribution can be obtained from Eq.
(\ref{e8}) by replacing one of the propagators 
\bea
\label{e13}
{S_q}_{\alpha\beta}^{ab} \rar - \frac{1}{4} \bar q^a A_j q^b
(A_j)_{\alpha\beta}~,
\eea
where $A_j = \left\{
1,~\gamma_5,~\gamma_\alpha,~i\gamma_5\gamma_\alpha,
~\sigma_{\alpha\beta}/\sqrt{2}\right\}$ is the full set of Dirac matrices
and sum over $A_j$ is implied, and the other two propagators are the full
ones.

It follows from Eqs. (\ref{e8}) and (\ref{e9}) that in order to calculate
the nonperturbative contributions to the theoretical part of
the correlator functions, one needs to know the matrix elements 
$\lla \gamma \vel \bar q A_j q \ver 0 \rra$ of the nonlocal operators between
photon and vacuum states. Up to twist--4, non--zero matrix elements of the
nonlocal operators are determined in terms of the photon wave functions as
follows\cite{R19,R20}:
\bea
\label{e14}
\la \gamma (q) \ve \bar q \gamma_\alpha \gamma_5 q \ve 0 \ra \es \frac{f}{4}
e_q \epsilon_{\alpha \beta \rho \sigma} \varepsilon^\beta
q^\rho x^\sigma \int_0^1 du \, e^{i u qx} \psi(u)~, \nnb \\
\la \gamma (q) \ve \bar q \sigma_{\alpha \beta} q \ve 0 \ra \es
i e_q \qq \int_0^1 du \, e^{i u q x} \Bigg\{ (\varepsilon_\alpha q_\beta -
\varepsilon_\beta q_\alpha) \Big[ \chi \phi(u) + x^2 \Big(g_1(u) -
g_2(u)\Big)
\Big]  \nnb \\
\ar \Big[ qx (\varepsilon_\alpha x_\beta - \varepsilon_\beta x_\alpha) +
\varepsilon x (x_\alpha q_\beta - x_\beta q_\alpha) \Big] g_2
(u) \Bigg\}~,
\eea
where $\chi$ is the magnetic susceptibility of the quark condensate,
$e_q$ is the quark charge, $\phi(u)$ and $\psi(u)$
are the leading twist--2 photon wave functions, while $g_1(u)$ and $g_2(u)$
are the twist--4 functions, respectively. Note that twist--3 photon wave
functions are neglected in further calculations since their contribution
changes the results about 5\%.
 
Using Eqs. (\ref{e11}) and (\ref{e14}), we can calculate theoretical part of
the correlator functions for $\Lambda$ and $\Sigma^+$ baryons from 
Eqs. (\ref{e8}) and (\ref{e9}). The sum rule is obtained by equating the 
theoretical and phenomenological parts of the corresponding correlator 
function. Performing double Borel transformations on the variables $p_1^2=p^2$ 
and $p_2^2=(p+q)^2$ in order to suppress the continuum and higher state 
contributions (for a discussion concerning this point, see 
\cite{R21}--\cite{R23}), the following sum rules for the $\Lambda$ and 
$\Sigma^+$ baryons are obtained:
\bea
\label{e15}
\lefteqn{
\lambda^2_\Lambda \mu_\Lambda e^{-M^2/M_\Lambda^2} =}\nnb \\
&&\frac{1}{192 \pi^4} M^6 E_2(x) \Big[(-1+t^2) (e_u+e_d) + (13+10 t + 13 t^2)
e_s \Big] \nnb \\
\ek \frac{m_s}{48 \pi^2} M^4 E_1(x) (-5+4 t + t^2)\chi \varphi(u_0) \qq
(e_u+e_d) \nnb \\
\ek \frac{1}{96 \pi^2} M^4 E_1(x) f \psi (u_0) \Big[(-1+t)^2 (e_u+e_d) +
(13+10 t + 13 t^2)e_s \Big] \nnb \\
\ar \frac{m_s}{12 \pi^2} M^2 E_0(x) (-1-4 t +5 t^2) \qq 
\ga \gamma_E - \ln \frac{M^2}{\Lambda^2} \dr e_s \nnb \\
\ek \frac{m_s}{288 \pi^2} m_0^2 \qq \Big[ 9(-1+t^2) (e_u+e_d) + 
(-3-12 t + 16 t^2) e_s\Big] \ga \gamma_E - \ln \frac{M^2}{\Lambda^2} 
\dr\nnb \\
\ek \frac{2}{9} \Big[ g_1 (u_0) - g_2 (u_0)\Big] \qq \Big[ (-5 + 4t + t^2) \qs
+ (-1+t)^2 \qq \Big] (e_u+e_d)\nnb\\
\ar \frac{m_s}{6 \pi^2} (-5 + 4t + t^2) \Big[ g_1 (u_0) - g_2 (u_0)\Big]
M^2 E_0 (x) \qq (e_u+e_d)\nnb \\
\ar\frac{4}{9} (-1 - 4t - 5t^2) \Big[ g_1 (u_0) - g_2 (u_0)\Big] 
\qs \qu e_s \nnb \\
\ar \frac{m_s}{48 \pi^2} M^2 E_0(x) \Big[(-1+t)^2 \qs - 2 (-5+ 4t +t^2 ) \qu
\Big] (e_u+e_d)\nnb \\
\ar \frac{m_s}{12 \pi^2} M^2 \qu (-1-4t+5t^2) e_s \nnb \\
\ar \frac{1}{18} M^2 E_0(x) \chi \varphi(u_0) \qq \Big[ (-5+4t+t^2) \qs +
(-1+t)^2 \qq \Big] (e_u+e_d)\nnb \\
\ar \frac{1}{9} M^2 E_0(x) (1+4t-5t^2)\chi \varphi(u_0) \qs \qq e_s \nnb \\
\ek \frac{m_s}{72}f \psi (u_0) \Big[ (-1+t)^2 \qs - 2(-5+4t+t^2) \qq
\Big] (e_u+e_d)\nnb \\
\ek \frac{1}{144} m_0^2 \chi \varphi (u_0) \qu \Big[ 2 (-5+4t+t^2) \qs +
(-3-2t+5t^2) \qq \Big] (e_u+e_d)\nnb \\
\ar\frac{1}{216} m_0^2 (-3 - 12t +16 t^2)\chi \varphi (u_0) 
\qs \qq e_s \nnb \\
\ar\frac{1}{18}\qq \Big[ (-5+4t+t^2) \qs (e_u+e_d) + (-11-2t+13t^2) \qq e_d
\big] \nnb \\
\ek \frac{m_s}{288\pi^2} m_0^2 \Big[(-1+t)^2 \qs -3 (-5+4t+t^2) \qq \Big]
(e_u+e_d)~, 
\eea 

\bea
\label{e16}
\lefteqn{
\lambda^2_{\Sigma^+} \mu_{\Sigma^+} e^{-M^2/M_{\Sigma^+}^2} =}\nnb \\
\ek \frac{1}{64 \pi^4} \Big[ M^6 E_2(x)- 2 M^4 E_1(x) f \psi (u_0) \Big]
\Big[(1+t)^2e_s -2(3+2t+3 t^2)e_u\Big]\nnb \\
\ar \frac{m_u}{4\pi^2} M^2 E_0(x) \Big[(-1+t^2)\qs + (-1+t)^2 \qu \Big] e_u
\ga \gamma_E - \ln \frac{M^2}{\Lambda^2} \dr\nnb \\
\ek \frac{m_s}{4\pi^2} M^2 E_0(x)(-1+t^2)\qu e_s \ga \gamma_E - \ln
\frac{M^2}{\Lambda^2} \dr\nnb \\
\ar \frac{m_s}{192\pi^2}m_0^2 \qu \Big[ (-15+14t^2) e_s - 12 (-3+t+3t^2) e_u
\Big] \ga \gamma_E - \ln \frac{M^2}{\Lambda^2} \dr\nnb \\
\ek\frac{m_u}{96\pi^2}m_0^2 \Big[(-7+19t^2) \qs + 3(1-3t+t^2)\qu \Big] e_u
\ga \gamma_E - \ln \frac{M^2}{\Lambda^2} \dr\nnb \\
\ek \frac{4}{3}(-1+t^2) \Big[ g_1 (u_0) - g_2 (u_0)\Big] \qs \qu e_s \nnb \\
\ar \frac{4}{3} \Big[ g_1 (u_0) - g_2 (u_0)\Big] \qu \Big[ (-1+t^2)\qs +
(-1+t)^2 \qu \Big] e_u\nnb \\
\ek \frac{1}{\pi^2} \qu \Big[ g_1 (u_0) - g_2 (u_0)\Big] M^2 E_0(x) 
\Big[(-1+t^2)m_s + (-1+t)^2 m_u \Big] e_u \nnb \\
\ar \frac{1}{\pi^2} (-1+t^2) \Big[ g_1 (u_0) - g_2 (u_0)\Big]  M^2 E_0(x) 
m_u \qs e_s \nnb \\
\ar \frac{1}{8\pi^2}M^4 E_1(x)\Big[(-1+t^2)m_s + (-1+t)^2 m_u \Big] \chi
\varphi(u_0) \qu e_u \nnb \\
\ek \frac{m_u}{8\pi^2} M^4 E_1(x) (-1+t^2) \chi \varphi(u_0) \qs e_s\nnb \\
\ar \frac{1}{3}(-1+t^2) M^2 E_0(x) \chi \varphi(u_0) \qs \qu e_s \nnb \\
\ek \frac{1}{3} M^2 E_0(x) \chi \varphi(u_0) \qu \Big[(-1+t^2)\qs + (-1+t)^2
\qu \Big] e_u \nnb \\
\ar\frac{m_u}{8\pi^2} M^2 E_0(x)\Big[ -4(-1+t^2) \qs + (5-2t+5t^2) \qu \Big]
e_u \nnb \\
\ar\frac{m_u}{8\pi^2} M^2 E_0(x) (1-6t+t^2) \qu e_s \nnb \\
\ar\frac{m_s}{8\pi^2}M^2 E_0(x) \Big[ (3+2t+3t^2) \qs - 6(-1+t^2)\qu \Big]
e_u \nnb \\
\ek\frac{m_s}{4\pi^2} M^2 E_0(x)(-1+t^2)\qu e_s \nnb \\
\ar \frac{m_u}{12} f \psi(u_0) \Big[6(-1+t^2) \qs - (3+2t+3t^2)\qu \Big]
e_u \nnb \\
\ek \frac{m_u}{12} f \psi(u_0) (1-6t+t^2) \qu e_s \nnb \\
\ek \frac{m_s}{12} f \psi(u_0) \Big[(3+2t+3t^2) \qs -
6(-1+t^2)\qu\Big]e_u\nnb \\
\ek \frac{1}{6}\qu \Big[(-1+t)^2 \qu e_s -6(-1+t^2)\qs e_u\Big]\nnb \\
\ar \frac{1}{72} m_0^2 \chi \varphi(u_0) \qu \Big[ (-1+t^2) \qs +
3(1-3t+t^2)\qu\Big] e_u\nnb \\
\ar \frac{1}{144} m_0^2 (15-14t^2) \chi \varphi(u_0) \qs \qu e_s\nnb \\
\ar\frac{m_u}{48\pi^2} m_0^2 \Big[9(-1+t^2)\qs - (3+2t+3t^2)\qu
\Big]e_u \nnb \\
\ek\frac{m_u}{24\pi^2}m_0^2 (1-4t+t^2)\qu e_s\nnb \\
\ek\frac{m_s}{48\pi^2}m_0^2 \Big[ (3+2t+3t^2) \qs-9(-1+t^2)\qu \Big] e_u~.  
\eea
In Eqs. (\ref{e15}) and (\ref{e16}) the functions
\bea
E_n(x) = 1 - e^x \sum_{k=0}^n \frac{x^k}{k\!}~,\nnb
\eea
are used to subtract the continuum and higher state contributions,  
$x=s_0/M^2$ and $s_0$ is the continuum threshold. Moreover
\bea
u_0 \es \frac{M_2^2}{M_1^2+M_2^2}~,\nnb \\
M^2 \es  \frac{M_1^2 M_2^2}{M_1^2+M_2^2}~,\nnb 
\eea
but, since we are dealing with just a single baryon, the Borel
parameters $M_1^2$ and $M_2^2$ should be set to equal each other, from which
it follows that $u_0=1/2$.
 
As has already been noted, the sum rules for the
$\Sigma^0,~\Sigma^-,~\Xi^0,~\Xi^-,~p,~n$ baryons can be obtained by making
the following replacements:
\bea
\mu_{\Sigma^0} \es \mu_{\Sigma^+} \ga m_{\Sigma^+} \rar m_{\Sigma^0},~
\lambda_{\Sigma^+} \rar \lambda_{\Sigma^0},~e_u \rar
\frac{e_u+e_d}{2},~s_0^{\Sigma^+} \rar s_0^{\Sigma^0}\dr~,\nnb \\
\mu_{\Sigma^-} \es \mu_{\Sigma^+} \ga m_{\Sigma^+} \rar m_{\Sigma^-},~
\lambda_{\Sigma^+} \rar \lambda_{\Sigma^-},~e_u \rar e_d,~
s_0^{\Sigma^+} \rar s_0^{\Sigma^-}\dr~,\nnb \\
\mu_{\Xi^0} \es \mu_{\Sigma^+} \ga m_{\Sigma^+} \rar m_{\Xi^0},~
e_s \leftrightarrow e_u,~m_s \leftrightarrow m_u,~\qs \rar \qu ,~
s_0^{\Sigma^+} \rar s_0^{\Xi^0}\dr~,\nnb \\
\mu_{\Xi^-} \es \mu_{\Xi^0} \ga m_{\Xi^0} \rar m_{\Xi^-},~e_u
\leftrightarrow e_d,~ 
s_0^{\Xi^0} \leftrightarrow s_0^{\Xi^-}\dr~,\nnb \\
\mu_p \es \mu_{\Xi^-} \ga m_{\Xi^-} \rar m_p,~e_s \rar e_u,~m_s \rar m_u,~
\qs \rar \qu,~s_0^{\Xi^-} \rar s_0^p \dr~,\nnb \\
\mu_n \es \mu_p \ga e_u \rar\frac{e_u+e_d}{2},~s_0^p \rar s_0^n\dr~.\nnb 
\eea
Note that, after we make these replacements we set masses of $u$ and 
$d$ quarks to zero and assume $SU(2)$ flavor symmetry, which implies 
that $\qu=\qd$.

Obviously, it follows from Eqs. (\ref{e15}) and (\ref{e16}) that
one needs to know the residues $\lambda_B$ of the octet baryons in order 
to determine their magnetic moments. These residues are determined from mass
sum rules of the corresponding baryons:
\bea
\label{e18}
\lefteqn{
\lambda_{\Sigma^+}^2 e^{-m_{\Sigma^+}^2/M^2} =}\nnb \\
&&\frac{M^6}{256 \pi^4} E_2(x) (5 + 2 t + 5 t^2)
- \frac{m_0^2}{24 M^2} \qu \left[ 12 (-1 + t^2) \qs  +  (- 1 + t)^2 
\qu \right] \nnb \\
\ar \frac{m_s}{32 \pi^2} M^2 E_0(x)  \left[ (5 + 2 t + 5 t^2) \qs 
-12 (-1 + t^2) \qu \right] \nnb \\
\ek \frac{m_0^2}{96 \pi^2} m_s \left[ (5 + 2 t + 5 t^2) \qs 
- 3 (-1 +  t^2) \qu \right] \\
\ar \frac{\qu}{6} \left[ 6 (-1 + t^2) \qs + (-1 + t)^2 \qu \right]\nnb \\
\ek \frac{3 m_s}{16 \pi^2} m_0^2 \qu (-1 + t^2) 
\left( \gamma_E - \ln \frac{M^2}{\Lambda^2} \right)\nnb
\\ \nnb \\ \nnb
\label{e20}
\lefteqn{
\lambda_{\Xi^-}^2 e^{-m_{\Xi^-}^2/M^2} =} \nnb \\
&&\frac{M^6}{256 \pi^4} E_2(x) (5 + 2 t + 5 t^2)
- \frac{m_0^2}{24 M^2} \qs \left[ 12 (-1+t^2) \qd  
+ (- 1 + t)^2 \qs \right] \nnb \\
\ar\frac{3 m_s}{16 \pi^2} M^2 E_0(x) \left[ -2 (-1 + t^2) \qd  
+ (1+t)^2\qs \right] \nnb \\
\ar \frac{\qs}{6} \left[ 6 (-1 + t^2) \qd + (-1 + t)^2 \qs \right] \\
\ek \frac{m_0^2 m_s}{96 \pi^2} \left[ 3 (-1 + t^2) \qd 
+ (7 + 10 t + 7 t^2) \qs \right] \nnb \\
\ek \frac{3 m_0^2 m_s}{16 \pi^2} \qd (-1 + t^2) \left( \gamma_E 
- \ln \frac{M^2}{\Lambda^2} \right)\nnb
\\ \nnb \\ \nnb
\label{e22}
\lefteqn{
\lambda_\Lambda^2 e^{-M_\Lambda^2/M^2} =} \nnb \\
&&\frac{1}{256 \pi^4} (5 + 2t + 5t^2) M^6 E_2(x) \nnb \\
\ar \frac{1}{72} (1 - t) \frac{m_0^2}{M^2}  \left\{ 4 (1 + 2 t) \qs 
\ga \qu + \qd \dr + (25 + 23 t) \qu \qd \right\}\nnb \\
\ar \frac{m_s}{96 \pi^2} M^2 E_0 (x) \left\{ 3 (5 + 2 t + 5 t^2 ) \qs + 2 (1 + 4 t 
- 5 t^2 ) \ga \qu + \qd \dr \right\} \nnb \\
\ar \frac{m_s}{32 \pi^2} m_0^2 \ga \qu + \qd \dr (1 - t^2 ) \left\{\gamma_E 
- \ln \left(\frac{M^2}{\Lambda^2} \right) \right\} \\
\ek \frac{1}{18} (1 - t) \left\{  (1 + 5 t) \qs \ga \qu + \qd \dr 
+ (13 + 11 t) \qu \qd \right\} \nnb \\
\ek \frac{m_s}{192 \pi^2} m_0^2 \left\{ 2 (5 + 2 t + 5 t^2 ) \qs 
+ (-5 + 4 t + t^2 ) \ga \qu+\qd \dr \right\}\nnb
\\ \nnb \\ \nnb
\label{e24}
\lefteqn{
\lambda_{\Sigma^0}^2 e^{-M_{\Sigma^0}^2/M^2} =}\nnb \\
&&\frac{1}{256 \pi^4} (5 + 2 t + 5 t^2 ) M^6 E_2(x) \nnb \\
\ar \frac{m_s}{32 \pi^2} M^2 E_0(x) \left\{ (5 + 2 t + 5 t^2 ) \qs 
- 6 (-1 + t^2 ) \ga \qu+\qd \dr \right\} \nnb \\
\ar \frac{1}{24} \frac{m_0^2}{M^2} (1 - t)  \left\{ 6 (1 + t) \qs 
\ga \qu + \qd \dr + (-1 + t) \qu \qd \right\} \nnb \\
\ar \frac{3 m_s}{32 \pi^2} m_0^2 \ga \qu + \qd \dr (1 - t^2 ) \left\{\gamma_E 
- \ln \left(\frac{M^2}{\Lambda^2} \right) \right\} \\
\ek \frac{m_s}{192 \pi^2} m_0^2 \left\{ 2 (5 + 2 t + 5 t^2 ) \qs 
- 3 (-1 + t^2 ) \ga \qu+\qd \dr \right\} \nnb \\
\ek \frac{1}{6}  (1 - t) \left\{ 3 (1 + t) \qs \ga \qu + \qd \dr 
+ (-1 + t) \qu \qd \right\}\nnb
\eea


\section{Numerical Analysis}
In this section we present the numerical analysis of the sum rules for the
magnetic moments of the octet baryons. The main input parameters in the
LCQCD are the photon wave functions which have been introduced in the previous
section. In \cite{R18} and \cite{R19} it has been shown that the leading
photon wave functions do not deviate much from their asymptotic form.
Therefore in numerical calculations we shall use the following forms
of the photon wave functions \cite{R18,R20}
\bea
\phi(u) \es 6 u (1-u)~,~~~~~\psi(u) = 1~, \nnb \\
g_1(u) \es - \frac{1}{8}(1-u)(3-u)~,~~~~~g_2(u) = -\frac{1}{4} (1-u)^2~.\nnb
\eea
The values of the other input parameters that are used in the numerical
analysis are:
$f=0.028~GeV^2$, $\chi=-3.3~GeV^{-2}$ \cite{R24} (in \cite{R25} this
quantity is estimated to be $\chi=-4.4~GeV^{-2}$), 
$\qq(1 ~ GeV)=-(0.243)^3~GeV^3$, $m_0^2=(0.8\pm0.2)~GeV^2$ \cite{R26},
$m_s(1~GeV)=(150\pm 50)~MeV$, $\qs(1~GeV)=0.8 \qq(1~GeV)$ and
$\Lambda=0.5~GeV$. 
In the problem under consideration we have three auxiliary parameters,
namely parameter $t$ in the hadron interpolating currents, Borel mass square
$M^2$ and the continuum threshold $s_0$. No doubt, it is expected that any
physically measurable quantity must be independent of these auxiliary 
parameters. So, our problem is to find the appropriate regions, for which 
magnetic moments of octet baryons are independent of the above--mentioned 
parameters. 

For this aim we consider the following three--step procedure. In the first
step we attempt to find working region of $M^2$, where magnetic moments of
octet baryons are independent of the Borel parameter at fixed values of
$s_0$ and $t$. In Figs. (1)--(8) we present the dependence of the magnetic
moments of the corresponding baryons on $M^2$. From these figures we see
that the magnetic moments of octet baryons seem to be 
almost independent of $M^2$ for different choices of $t$ and $s_0$. However,
the working region for the Borel mass square parameter for the members of 
octet baryons are different, i.e., $0.9~GeV^2 <M^2<1.2~GeV^2$ for $p,~n$;
$1.3~GeV^2 <M^2<1.6~GeV^2$ for $\Sigma^-,~\Sigma^0,~\Sigma^+$ and $\Lambda$;
$1.7~GeV^2 <M^2<2.1~GeV^2$ for $\Xi^0$ and $\Xi^-$.

Before determining the magnetic moment of baryons, the next problem to be
considered is to find the working region of $t$, i.e. the region where the sum rules make sense.
For this purpose we consider the mass sum rules (Eqs.
(\ref{e18}--\ref{e24})). Two criteria should be satisfied for mass sum rules.
First of all, each mass sum rule 
must separately be positive. The second
criteria is that, the predicted mass of the baryons which is obtained by
taking the logarithmic derivative of the corresponding sum rule with respect to
$M^{-2}$
should be stable with respect to the variation of the parameter $t$. 
As a result of analysis of the mass sum rules we found that, the region
$-0.5<t<0.5$ is unphysical (see Fig. (9)).  
Our analysis shows that the working region of $t$ where the second
criteria is satisfied is given by $-0.6 < \cos\theta < 0.3$, where $\theta$
is defined through the relation $\theta=\tan^{-1} t$. This region of $\theta$
corresponds to $t < -1.4$ and $t>3.3$. We observe that the Ioffe current,
which corresponds to the choice $t=-1$ does not lie in the working
region.   

Our last attempt is
to determine a region for the parameter $t$ where the magnetic moments
of the octet baryons are independent of its variation. Here we stress
again that in the first step the working region for the Borel mass square
$M^2$ has been determined for which $\mu_B$ is practically independent of its
variation, as well as insensitive to the continuum threshold. 
We present in Fig (10) the dependence of the magnetic moments of the
octet baryons (in units of the nucleon magneton $\mu_N$) on $\cos\theta$, at
fixed values of $M^2$ (remember that the fixed values of $M^2$ are different
for each one of the octet baryons) and at fixed value of $s_0$ (again
different fixed values of $s_0$ are used for each baryon). From this figure
we observe that $\mu_B$ is practically independent of the parameter $t$ in
the region $-0.6 < \cos\theta <0.3$ and insensitive to 
the variation of the
continuum threshold $s_0$. Our final results on the magnetic moments of the
octet baryons are presented in Table (1) . 
For completeness, in this table
we also present the predictions of other approaches on the magnetic moments
of octet baryons and the experimental results as well as our results for the
value of the magnetic susceptibility $\chi = -4.4~GeV^2$. It follows from our
analysis that light cone QCD sum rule prediction on magnetic moments of
octet baryons are close to their experimental values for $\chi =
-3.3~GeV^2$, but considerable departure is observed for $\chi = -4.4~GeV^2$.
Obviously, it follows from Table 1 that similar situation takes place for the
traditional sum rule predictions on the magnetic moments of octet baryons.
The uncertainties in the results we have presented could be attributed to the
variation of $s_0$, Borel parameter $M^2$, twist--3 photon wave functions
which are neglected and errors in the input parameters such as $\qq$,
$m_0^2$ and $m_s$. All errors are added quadratically to the final
predictions. It should be stressed here that the predictions on magnetic moments at
$t=-1$, which corresponds to the Ioffe current, are not reliable.

From our results we see that the values of $\mu_{\Sigma^+} \approx \mu_p$;
$2 \mu_\Lambda \approx \mu_n$; $\mu_{\Xi^0}\approx \mu_n$ and $\mu_{\Xi^-}\approx
\mu_{\Sigma^-}$. It should be added here that, in exact $SU(3)$ symmetry,
the relations between the magnetic moments are as follows \cite{R27}:
\bea       
\label{e25}
\mu_{\Sigma^+} \es \mu_p~,\nnb \\
2 \mu_\Lambda \es \mu_n~,\nnb \\
\mu_{\Sigma^-}+ \mu_n  \es - \mu_p~,\nnb \\
\mu_{\Xi^-} \es \mu_{\Sigma^-}~,\nnb \\
\mu_{\Xi^0} \es \mu_n~.
\eea
Violation to the relations (\ref{e25}) comes from $SU(3)$ breaking terms
(the mass of the $s$ quark, $s$ quark condensate, etc.). So, our predictions
give us a clue about sizable $SU(3)$ symmetry breaking effects.

In summary, we have calculated magnetic moments of the octet baryons in
light cone QCD sum rules, using the general form of the baryonic
interpolating currents. We have obtained that light cone QCD sum rule
predictions on magnetic moments of the octet baryons are in good agreement
with the experimental data at $\chi = -3.3~GeV^2$.

\section*{Acknowledgements}
We would like to thank V. S. Zamiralov for fruitful discussions.

Note added: After we have completed the present work, we have received
a recent work \cite{R33} where the mass sum rules for general form of
the interpolating currents for the baryons are calculated. Our results on mass
sum rules coincide with with the ones predicted in this work. 

\newpage

\landscape
\newcommand{\rb}[1]{\raisebox{1.5ex}[0pt]{#1}}

\begin{table}[ht]
\renewcommand{\arraystretch}{1.3}
\addtolength{\arraycolsep}{1.5pt}
$$
\begin{array}{|l|c|c|c|c|c|c|c|c|c|c|}
\hline \hline
      &    &    & 
\multicolumn{2}{|c|}{\mbox{\rm QCDSR}} &    &    &    & 
\multicolumn{2}{|c|}{\mbox{\rm LCQSR}} &  \\
\cline{4-5} \cline{9-10}                    
                    &  \rb{\mbox{\rm NQM}}   & \rb{\mbox{\rm SQM}}
& \chi=-3.3   & \chi=-4.4  &   \rb{\mbox{\rm QCDSA}}  & \rb{$\chi$\mbox{\rm PT}}
& \rb{\mbox{\rm SKRM}}  					 	 & \chi=-3.3    & \chi=-4.4   & \rb{\mbox{\rm EXP}} \\ \hline         
\mu_p          &  2.87 &  2.75 &  2.72 &  3.55 & 2.54  &  2.793  &  2.36 &  2.7 \pm 0.5 &  3.5 \pm 0.5 	&  2.79         \\ \hline
\mu_n          & -1.91 & -1.84 & -1.65 & -2.06 & -1.69 & -1.913  & -1.87 & -1.8 \pm 0.35& -2.3 \pm 0.4	& -1.91         \\ \hline
\mu_{\Sigma^+} &  2.62 &  2.65 &  2.52 &  3.30 & 2.48  &  2.458  &  2.46 &  2.2 \pm 0.4	&  2.9 \pm 0.4	&  2.46\pm 0.01 \\ \hline
\mu_{\Sigma^-} & -1.20 & -1.02 & -1.13 & -1.38 & -0.90 & -1.16   & -1.16 & -0.8 \pm 0.2	& -1.1 \pm 0.3	& -1.16\pm 0.03 \\ \hline
\mu_{\Xi^0}    & -0.63 & -1.44 & -0.89 & -0.98 & -1.49 & -1.25   & -1.25 & -1.3 \pm 0.3	& -1.3 \pm 0.4	& -1.25\pm 0.01 \\ \hline
\mu_{\Xi^-}    & -1.44 & -0.52 & -1.18 & -1.27 & -0.63 & -0.6531 & -0.65 & -0.7 \pm 0.2	& -1.0 \pm 0.2	& -0.65         \\ \hline
\mu_{\Lambda}  & -0.63 & -0.67 & -0.50 & -0.80 & -0.69 & -0.613  & -0.60 & -0.7 \pm 0.2	& -0.9 \pm 0.2	& -0.61         \\ \hline \hline
\end{array}
$$
\caption{ Predictions of various approaches for the octet baryon
magnetic moments: naive quark model (NQM, see ref. in \cite{R31});
static quark model (SQM) \cite{R32}; QCD sum rules (QCDSR) \cite{R9};
QCD string approach (QCDSA) \cite{R28}; chiral perturbation theory
($\chi$PT) \cite{R29}; skyrme model (SKRM) \cite{R30};
present work (LCQSR). For completeness we present the experimental values of the
octet baryons. All the values in the table are given in units of nuclear
magneton $\mu_N$.}

\renewcommand{\arraystretch}{1}
\addtolength{\arraycolsep}{-5pt}
\end{table}

\endlandscape

\newpage

\section*{Figure captions}
{\bf Fig. (1)} The dependence of the magnetic moment (in units of
nuclear magneton $\mu_N$) of $p$ on the Borel parameter $M^2$, at
the continuum threshold $s_0=2.0~GeV^2$.\\ \\
{\bf Fig. (2)} The same as in Fig. (1), but for $n$.\\ \\
{\bf Fig. (3)} The same as in Fig. (1), but for $\Lambda$, at 
the continuum threshold $s_0=2.5~GeV^2$ and $m_s=0.15~GeV$.\\ \\
{\bf Fig. (4)} The same as in Fig. (3), but for $\Sigma^0$, at       
the continuum threshold $s_0=3.0~GeV^2$ and $m_s=0.15~GeV$.\\ \\
{\bf Fig. (5)} The same as in Fig. (4), but for $\Sigma^-$.\\ \\
{\bf Fig. (6)} The same as in Fig. (4), but for $\Sigma^+$.\\ \\
{\bf Fig. (7)} The same as in Fig. (3), but for $\Xi^0$, at       
the continuum threshold $s_0=3.2~GeV^2$ and $m_s=0.15~GeV$.\\ \\
{\bf Fig. (8)} The same as in Fig. (7), but for $\Xi^-$.\\ \\
{\bf Fig. (9)} The dependence of $\lambda_B^2$  on $t$, for all 
members of the octet baryons, at the continuum threshold
$s_0=2.0~GeV^2$ (for $p$ and $n$), $s_0=2.5~GeV^2$ 
(for $\Lambda$),  $s_0=3.0~GeV^2$ (for $\Sigma$) and 
$s_0=3.2~GeV^2$ (for $\Xi$).\\ \\
{\bf Fig. (10)} The dependence of the magnetic moments of the octet 
baryons on $\cos\theta$ at the continuum 
threshold $s_0=2.0~GeV^2$ (for $p$ and $n$), $s_0=2.5~GeV^2$ 
(for $\Lambda$),  $s_0=3.0~GeV^2$ (for $\Sigma$) and 
$s_0=3.2~GeV^2$ (for $\Xi$), and at $M^2=1~GeV^2$ (for $p$ and $n$), 
$M^2=1.5~GeV^2$ (for $\Sigma$ and $\Lambda$) and $M^2=1.9~GeV^2$ 
(for $\Xi$).  
\begin{figure}[hbtp]
\begin{center}
  \includegraphics[height=12cm, angle=-90]{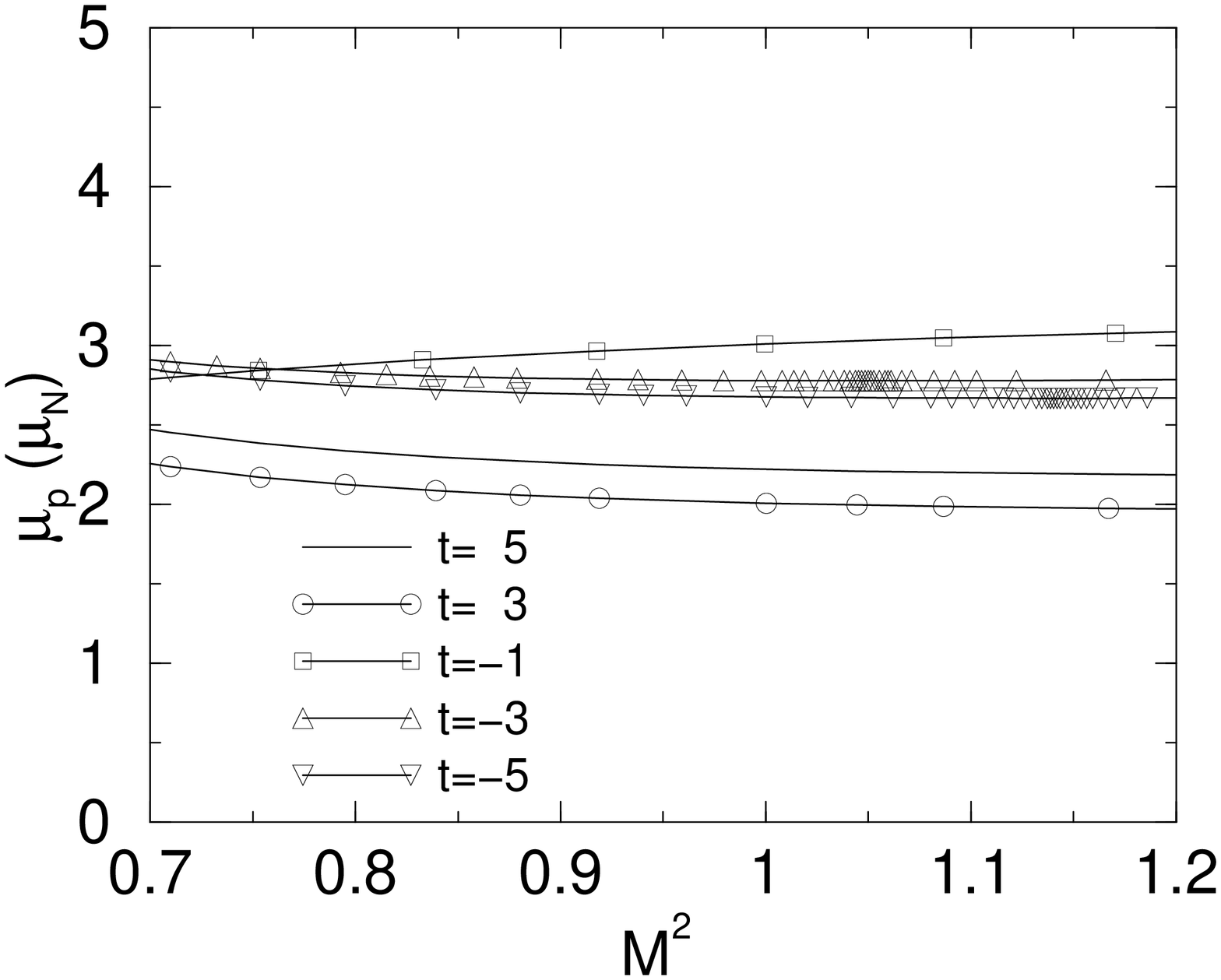}
\end{center}
\caption{}
\end{figure}

\begin{figure}[hbtp]
\begin{center}
  \includegraphics[height=12cm, angle=-90]{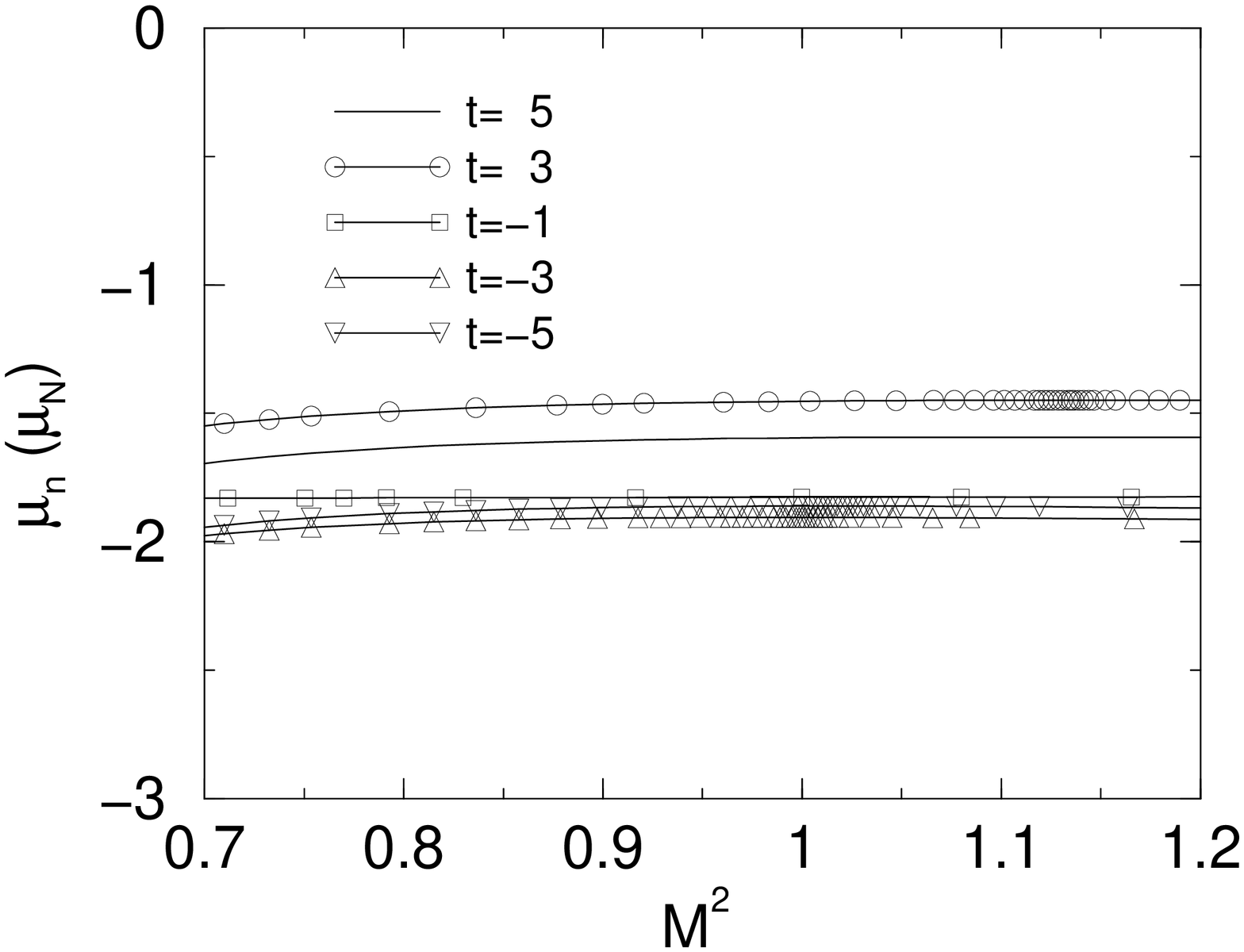}
\end{center}
\caption{}
\end{figure}

\begin{figure}[hbtp]
\begin{center}
  \includegraphics[height=12cm, angle=-90]{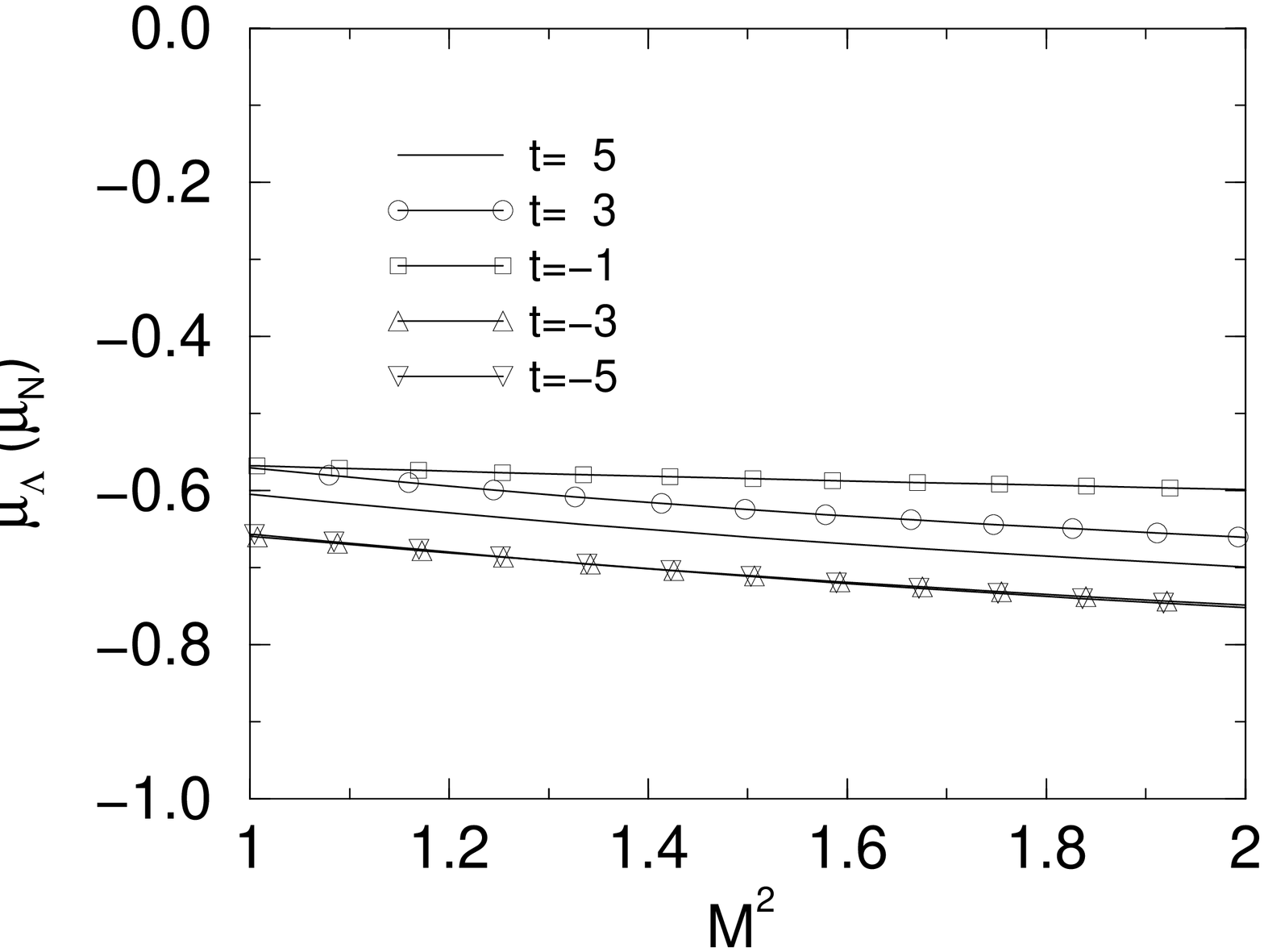}
\end{center}
\caption{}
\end{figure}

\begin{figure}[hbtp]
\begin{center}
  \includegraphics[height=12cm, angle=-90]{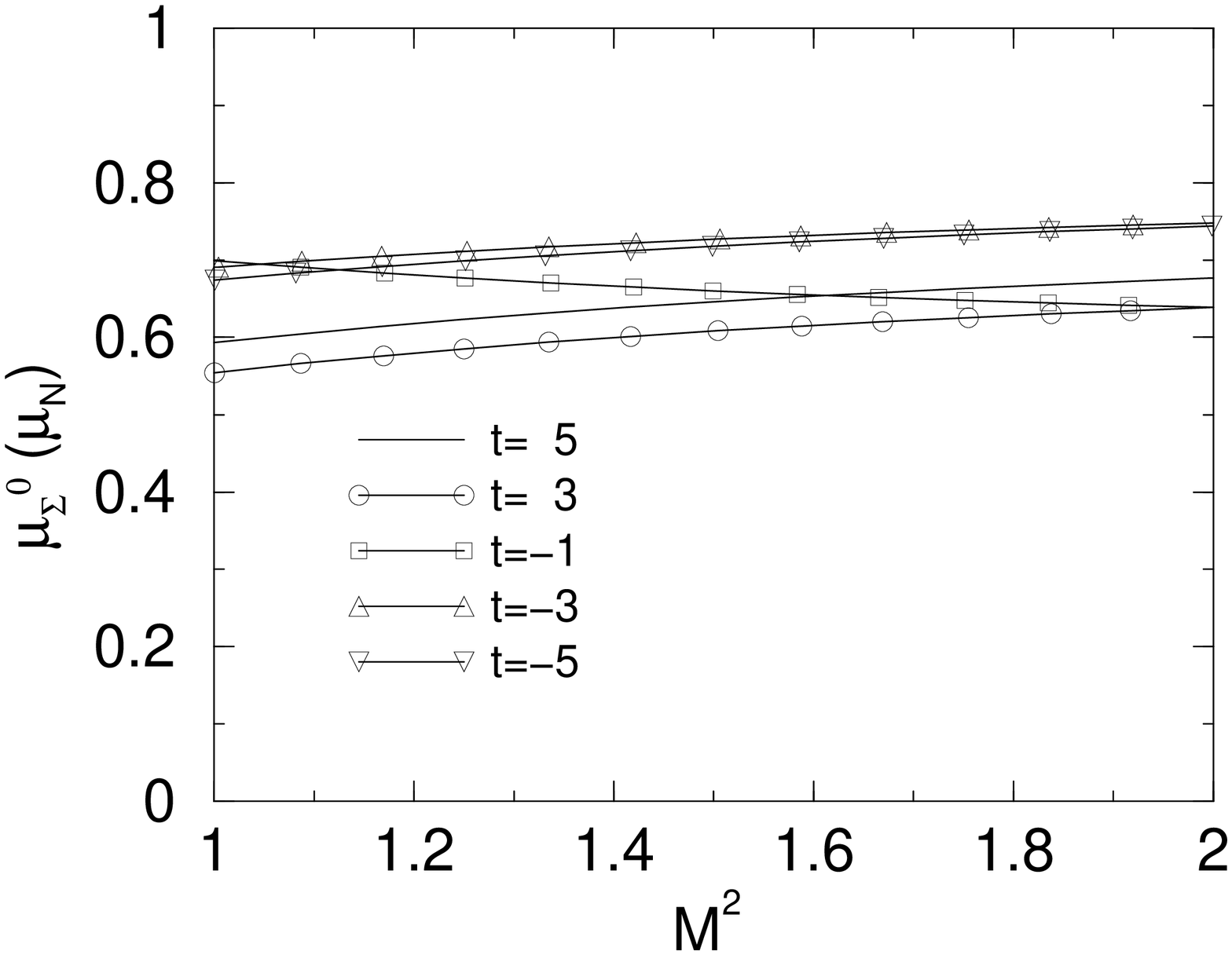}
\end{center}
\caption{}
\end{figure}

\begin{figure}[hbtp]
\begin{center}
  \includegraphics[height=12cm, angle=-90]{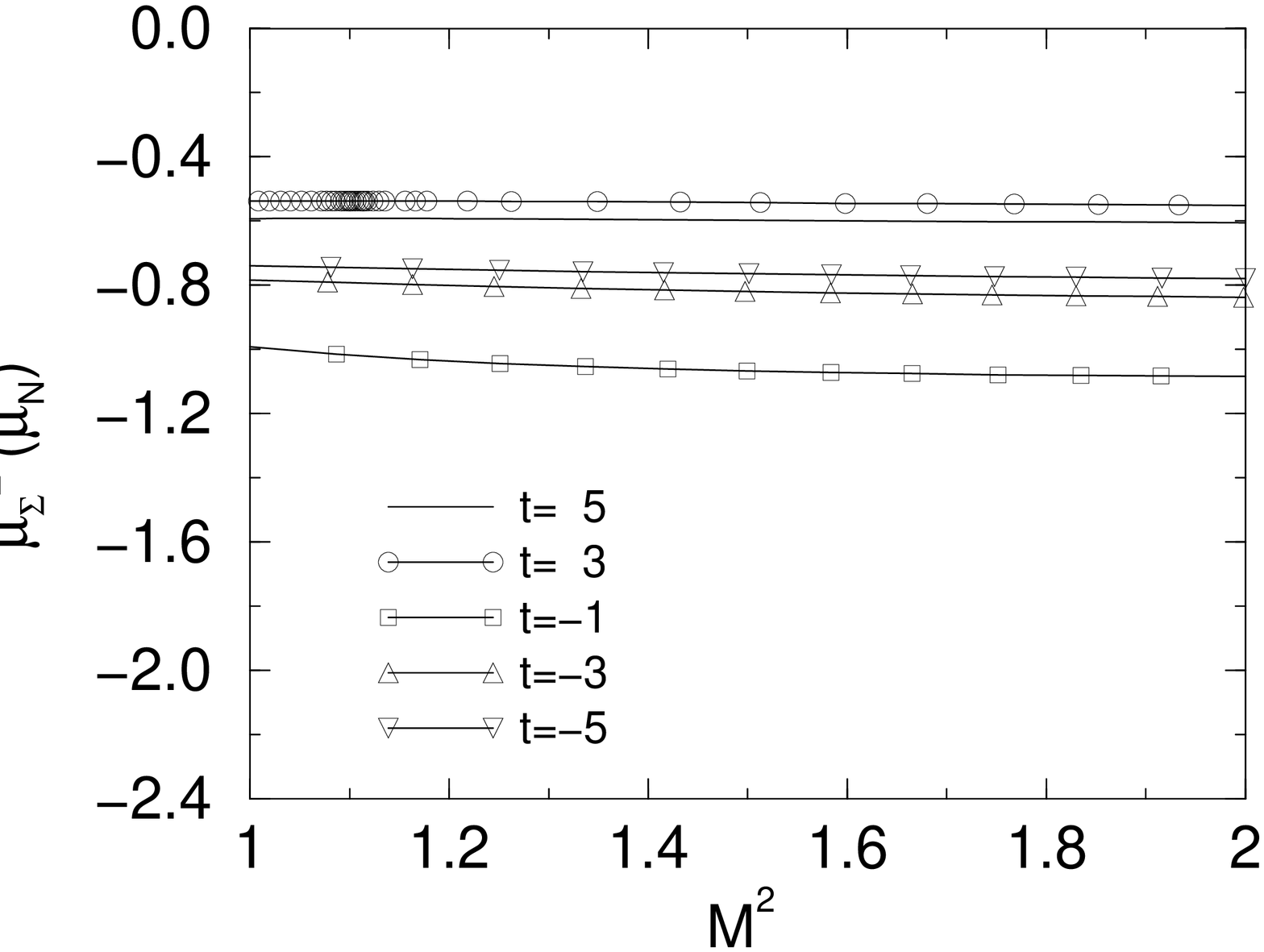}
\end{center}
\caption{}
\end{figure}

\begin{figure}[hbtp]
\begin{center}
  \includegraphics[height=12cm, angle=-90]{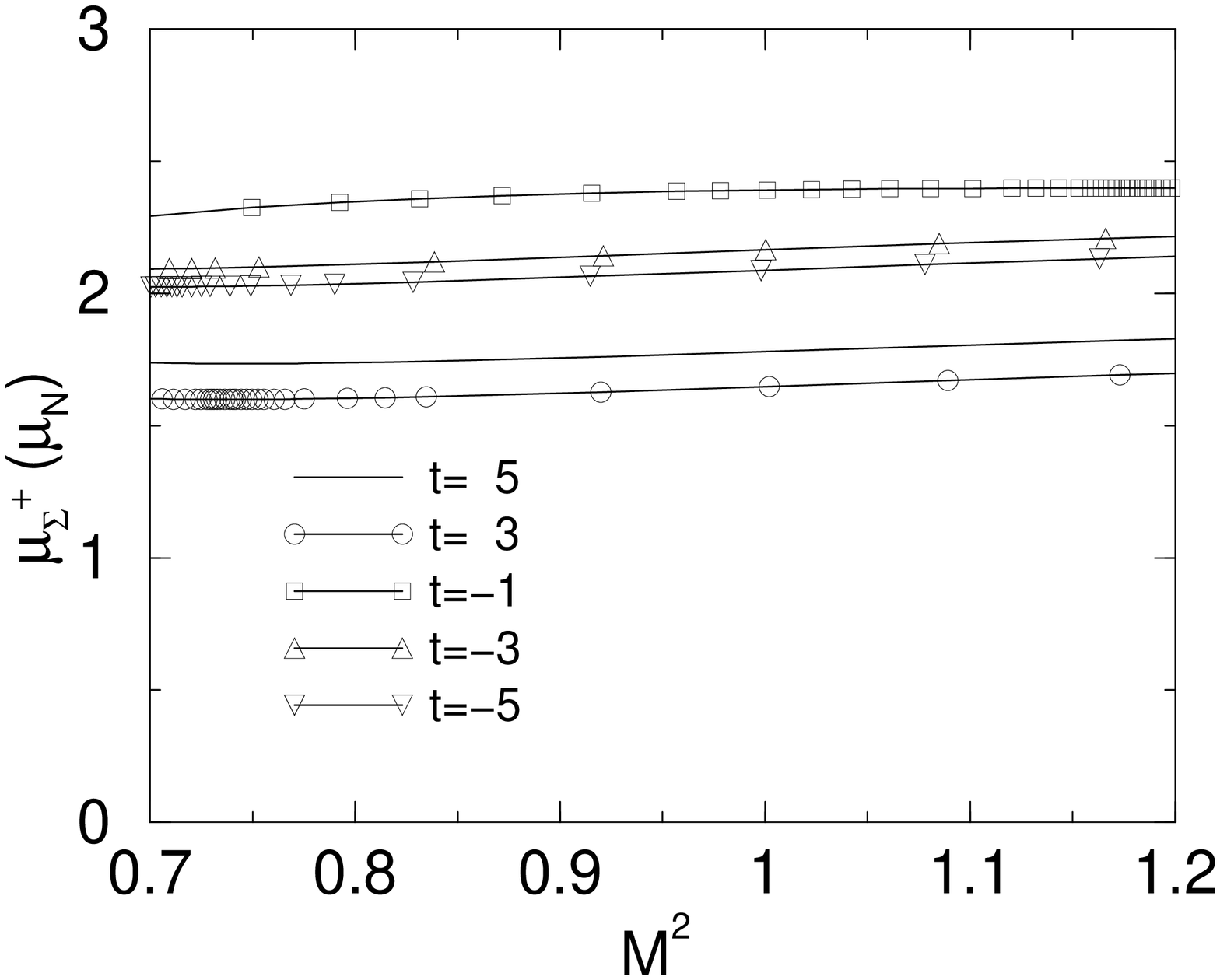}
\end{center}
\caption{}
\end{figure}

\begin{figure}[hbtp]
\begin{center}
  \includegraphics[height=12cm, angle=-90]{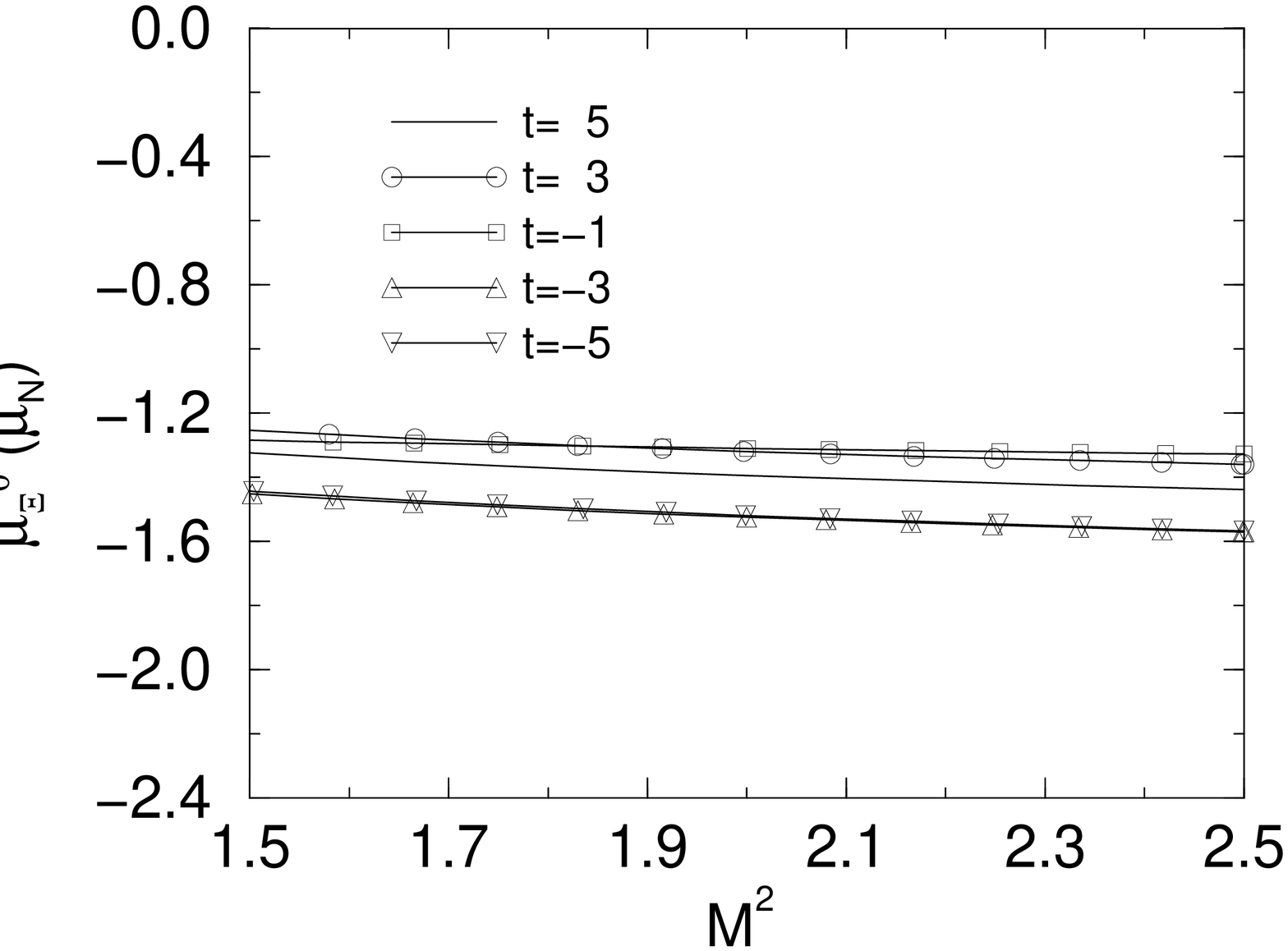}
\end{center}
\caption{}
\end{figure}

\begin{figure}[hbtp]
\begin{center}
  \includegraphics[height=12cm, angle=-90]{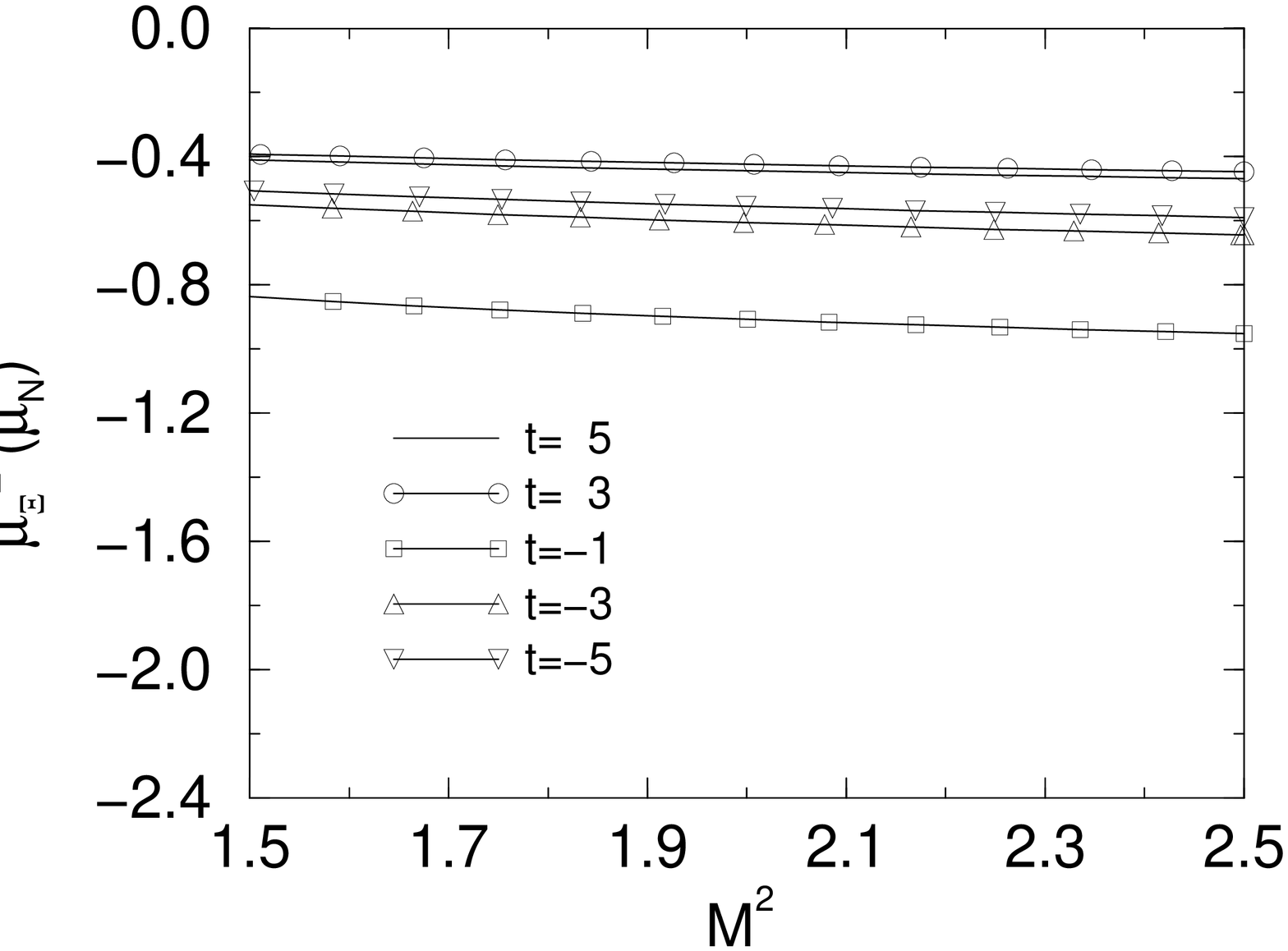}
\end{center}
\caption{}
\end{figure}

\begin{figure}[hbtp]
\begin{center}
  \includegraphics[height=12cm, angle=-90]{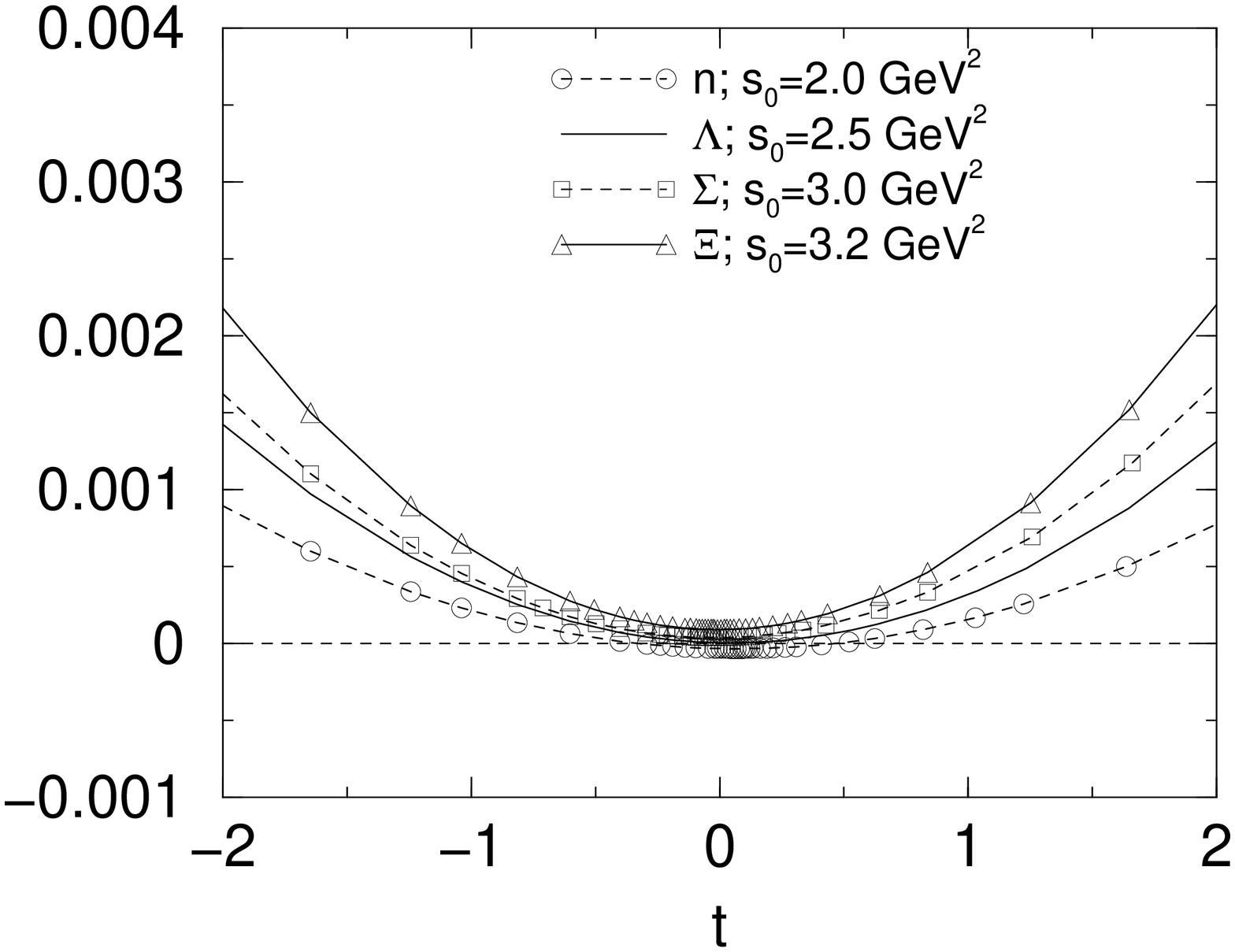}
\end{center}
\caption{}
\end{figure}

\begin{figure}[hbtp]
\begin{center}
  \includegraphics[height=12cm, angle=-90]{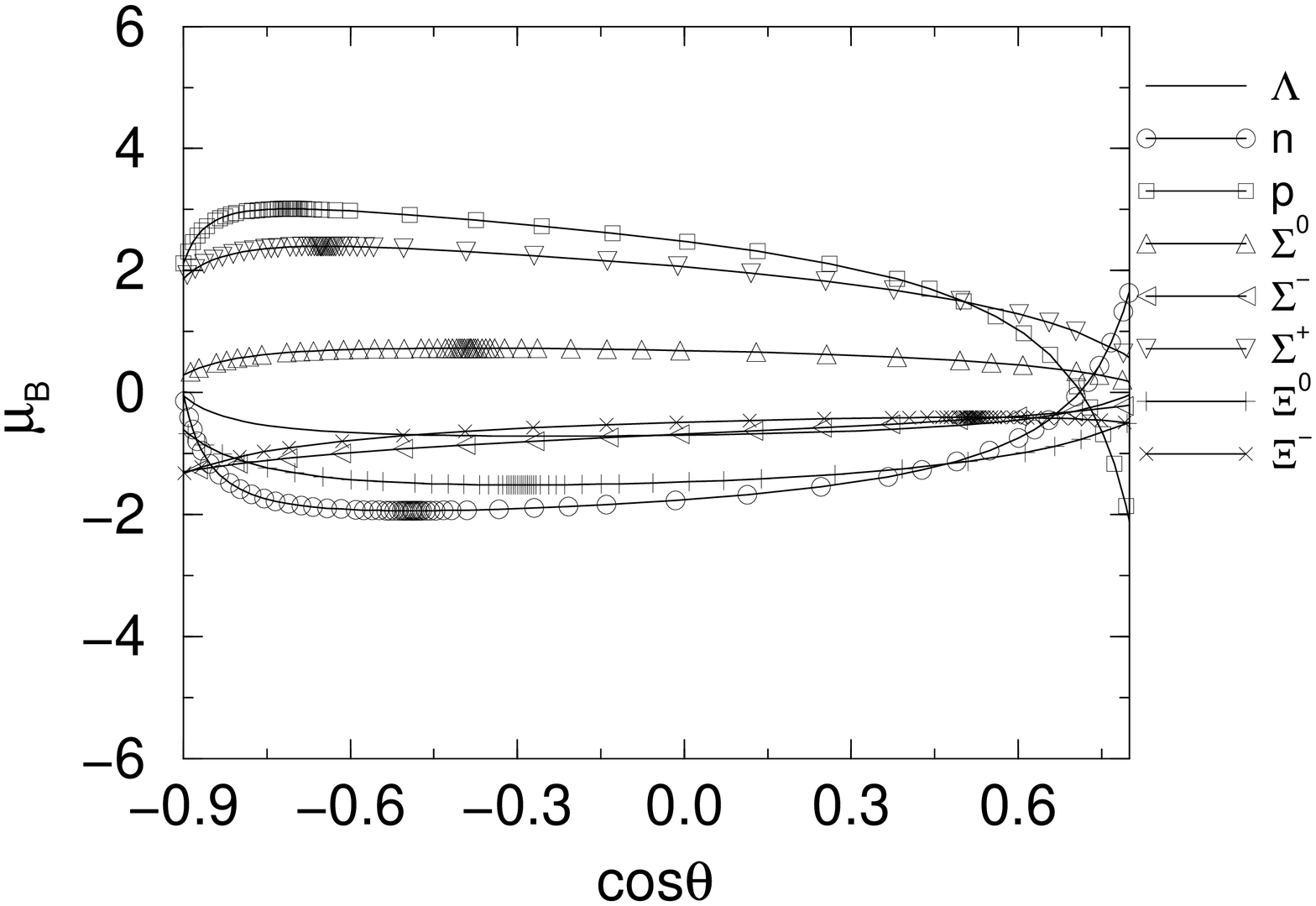}
\end{center}
\caption{}
\end{figure}

\end{document}